# Motion in time of a swinging Atwood's machine

Giorgio Carizi

Saipem S.p.A., Fano Engineering Center; Via Toniolo 1, 61032 Fano (PU), Italy

## Abstract

An exact solution for the motion in time of the swinging Atwood's machine is presented. The trajectories of the so called "teardrop-heart" orbits are obtained considering an extended Hamiltonian and a time transformation first introduced by Poincaré. The motion is isochronous with respect to the launch angle.

## 1. INTRODUCTION

A swinging Atwood's machine (SAM), Figure 1, is an ordinary Atwood machine in which, however, one of the weights can swing in a vertical plane [1].

In the polar coordinates (R, θ) the Lagrangian of this conservative dynamical system takes the form

$$\mathcal{L} = \frac{1}{2}(m + M)\dot{R}^2 + \frac{1}{2}mR^2\dot{\theta}^2 - gR(M - m\,cos\theta)$$

where M is the mass of the non-swinging bob, m is the mass of the swinging bob, g is the gravitational acceleration, θ is the polar angle measured from the negative vertical y-axis as shown in Figure 1 and the dot means derivative with respect to the (real) time t.

Setting the length of the string ($\ell$) as the unit length, $\sqrt{\ell/g}$ as the unit of time, $mg\ell$ as the unit of energy and introducing the mass ratio $\mu = M/m$, the dimensionless Lagrangian takes the form

$$L = \frac{1}{2}(1+\mu)\dot{r}^2 + \frac{1}{2}r^2\dot{\theta}^2 - r(\mu - cos\theta) \quad (1)$$

where $\mathcal{L} = (mg\ell)L$, $R = \ell r$, $t = \sqrt{\ell/g}\,\varpi$ and here dot means derivative with respect to the dimensionless time $\varpi$.

Tufillaro [2] showed that, when $\mu = 3$, by the change of variables

$$r = \frac{1}{2}(\xi^2 + \eta^2) \quad (2)$$

$$\theta = 2\,arctan[(\xi^2 - \eta^2)/2\xi\eta] \quad (3)$$

it's possible to obtain the exact explicit orbit equation for the mass m which start at the origin with initial conditions $(r_0 = \varepsilon, \dot{r}_0, \theta_0, \dot{\theta}_0)$, where ε is an infinitesimal quantity. The orbit is a symmetrical loop and return to the origin, no matter what the launch angle or speed.

Tufillaro [2] obtained the exact solution using the Hamilton-Jacobi theory and called these orbits "teardrop-heart" orbits.

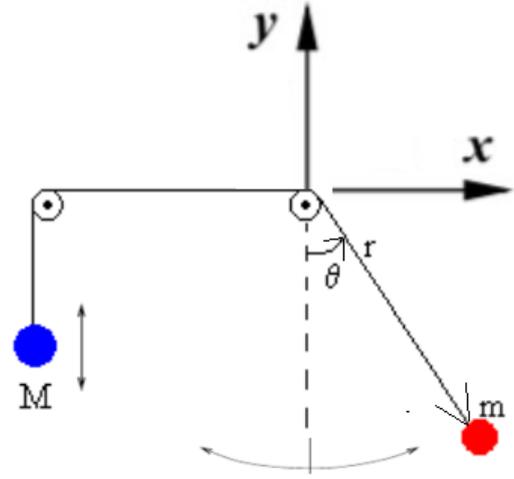

Figure 1. Swinging Atwood machine

In the $(\xi, \eta)$ variables the Hamiltonian of the system takes the form

$$H = \frac{1}{(\xi^2+\eta^2)}\left[\frac{1}{8}\left(p_\xi^2 + p_\eta^2\right) + 2(\xi^4 + \eta^4)\right] \quad (4)$$

where

$$p_\xi = 4\dot{\xi}(\xi^2 + \eta^2) \qquad p_\eta = 4\dot{\eta}(\xi^2 + \eta^2) \quad (5)$$

In the new variables the initial conditions read

$$\xi_0 = \sqrt{r_0}\sqrt{1 + sin(\theta_0/2)} \qquad (p_\xi)_0 = 4\dot{\xi}_0(\xi_0^2 + \eta_0^2)$$

$$\eta_0 = \pm\sqrt{r_0}\sqrt{1 - sin(\theta_0/2)} \qquad (p_\eta)_0 = \pm 4\dot{\eta}_0(\xi_0^2 + \eta_0^2) \quad (6)$$

where

$$\dot{\xi}_0 = \frac{\xi_0 \dot{r}_0}{2r_0} + \frac{r_0 \dot{\theta}_0}{4\xi_0}cos\left(\frac{\theta_0}{2}\right) \text{ and } \dot{\eta}_0 = \frac{\eta_0 \dot{r}_0}{2r_0} - \frac{r_0 \dot{\theta}_0}{4\eta_0}cos\left(\frac{\theta_0}{2}\right).$$

Hamilton-Jacobi theory provided two equations, the orbit equation (Eq. (16) in [2]) and the trajectory equation (Eq. (17) in [2]). Tufillaro successfully solved the orbit equation, but no additional details were provided of the trajectory equation.

## 2. THE REGULARIZING TRANSFORMATION

Following Poincaré ([3], [4] p.35, [5], [6]) a new time variable (τ) and a new Hamiltonian ($\mathcal{H}$) are considered: τ (often called the fictitious time) is connected to the real time $\varpi$ through the differential relation

$$d\varpi = (\xi^2 + \eta^2)d\tau \quad (7)$$

while the new Hamiltonian is as follows

$$\mathcal{H} = (\xi^2 + \eta^2)(H - E) \quad (8')$$

where E is the energy of the system. The complete expression for $\mathcal{H}$ read

$$\mathcal{H} = \frac{1}{8}(p_\xi^2 + p_\eta^2) + 2(\xi^4 + \eta^4) - E(\xi^2 + \eta^2) \quad (8)$$

The feature of Poincaré approach is to preserve the description of the system's dynamics: the equations of motion, in the time $\varpi$, generated by the Hamiltonian H (4) at a fixed value of the energy are equivalent to the equations of motion, in the time τ, generated by the Hamiltonian $\mathcal{H}$ (8) at the fixed "pseudo energy" equal to zero.

The Hamiltonian $\mathcal{H}$ (8), made up of two decoupled oscillators, can be separated into a new constant of motion ($I_2$)

$$\frac{1}{8}p_\xi^2 - E\xi^2 + 2\xi^4 = I_2 \quad (9)$$

$$\frac{1}{8}p_\eta^2 - E\eta^2 + 2\eta^4 = -I_2 \quad (10)$$

Plugging the initial conditions (6) into Eqs. (9) and (10) gives $I_2 = 0$ for all τ.

## 3. INTEGRATION OF THE EQUATIONS OF MOTION

The equations of motion in the fictitious time for the $(\xi, p_\xi)$ oscillator read

$$\frac{d\xi}{d\tau} = \frac{\partial \mathcal{H}}{\partial p_\xi} = \frac{1}{4} p_\xi \quad (11)$$

$$\frac{dp_\xi}{d\tau} = -\frac{\partial \mathcal{H}}{\partial \xi} = 2E\xi - 8\xi^3 \quad (12)$$

The constant of motion ($I_2$) is used to obtain from Eq. (9) $p_\xi$ as a function of ξ, which substituted in Eq. (11) make this last equation to be integrated straight away

$$-\frac{1}{2}\ln\left(\frac{1+\sqrt{1-\frac{2}{E}\xi^2}}{1-\sqrt{1-\frac{2}{E}\xi^2}}\right) = \frac{\sqrt{2E}}{2}\tau - \frac{1}{2}\ln(A) \quad (13')$$

$$\xi^2(\tau) = 2AE \frac{e^{-\sqrt{2E}\tau}}{\left(1+Ae^{-\sqrt{2E}\tau}\right)^2} \quad (13)$$

where A, obtained setting $\xi^2(\tau = 0) = \xi_0^2$, is very well approximate by the quantity $\frac{2E}{\xi_0^2}$.

Following the same steps, the motion in the fictitious time for the $(\eta, p_\eta)$ oscillator read

$$\eta^2(\tau) = 2BE \frac{e^{-\sqrt{2E}\tau}}{\left(1+Be^{-\sqrt{2E}\tau}\right)^2} \quad (14)$$

where B is very well approximate by the quantity $\frac{2E}{\eta_0^2}$.

The radius vector and the angle of displacement in the fictitious time of the swinging mass are obtained plugging Eqs. (13) and (14) in Eqs. (2) and (3), while the integration of Eq. (7) gives the relation between the fictitious and the real (dimensionless) time

$$\varpi = \sqrt{2E}[f(A) - g(A, \tau) + f(B) - g(B, \tau)] \quad (15)$$

where $f(\alpha) = \left(1 + \frac{1}{\alpha}\right)^{-1}$ and $g(\alpha, \tau) = \left(1 + \frac{e^{\sqrt{2E}\tau}}{\alpha}\right)^{-1}$. ($f(\alpha)$ has been obtained choosing the same origin for $\varpi$ and τ).

As τ goes to infinity, $g(\alpha, \tau) \to 0$ and $\varpi$ goes to the semi-period (T/2) of the oscillation

$$\lim_{\tau \to \infty} \varpi = 2\sqrt{2E} = \frac{T}{2} \quad (16)$$

In the "teardrop-heart" orbits the energy depends only on $\dot{r}_0$, i.e. $E = 2\dot{r}_0^2$, therefore, the trajectories are isochronous with respect to the $\theta$ variable: the period of the oscillation doesn't depend on the initial launch angle.

Figure 2 shows the trajectory of the swinging mass for initial conditions $(r = 10^{-8}, \dot{r}_0 = 0.3, \theta_0 = 3\pi/4, \dot{\theta}_0 = 0.1)$.

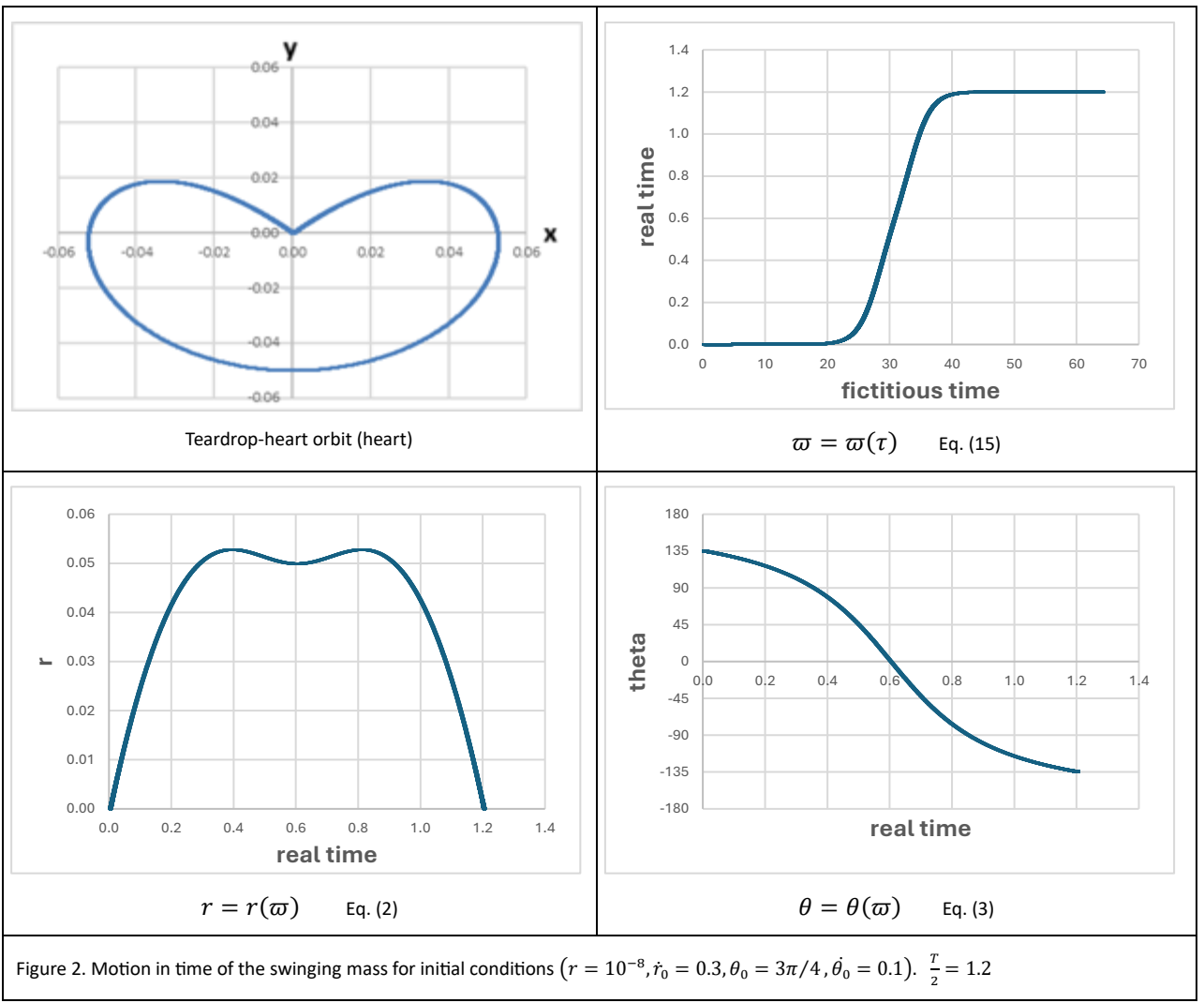

Figure 2. Motion in time of the swinging mass for initial conditions $(r = 10^{-8}, \dot{r}_0 = 0.3, \theta_0 = 3\pi/4, \dot{\theta}_0 = 0.1)$. $\frac{T}{2} = 1.2$